\documentclass[11pt]{article}
\usepackage[sc]{mathpazo}
\usepackage{fullpage}
\usepackage[authoryear,sectionbib,sort]{natbib}
\usepackage{graphicx}
\usepackage{adjustbox}
\usepackage[boxruled,linesnumbered]{algorithm2e}
\usepackage{multirow}
\usepackage{amsmath}
\usepackage[utf8]{inputenc}
\usepackage{lineno}
\usepackage{titlesec}
\usepackage{graphicx}
\usepackage{subcaption}
\usepackage{booktabs,multirow}
\usepackage{here}
\usepackage{url}

\titleformat{\section}[block]{\Large\bfseries\filcenter}{\thesection}{1em}{}
\titleformat{\subsection}[block]{\Large\itshape\filcenter}{\thesubsection}{1em}{}
\titleformat{\subsubsection}[block]{\large\itshape}{\thesubsubsection}{1em}{}
\titleformat{\paragraph}[runin]{\itshape}{\theparagraph}{1em}{}[. ]

\title{Understanding learning from EEG data: Combining machine learning and feature engineering based on hidden Markov models and mixed models}

\author{Gabriel R. Palma$^{1, 2, \ast}$ \and
Conor Thornberry$^{3}$ \and
Seán Commins$^{3}$ \and
Rafael A. Moral$^{1, 2}$}

\date{}

\begin{document}

\maketitle

\noindent{} 1. Hamilton Institute, Maynooth University, Maynooth, Ireland;

\noindent{} 2. Department of Mathematics and Statistics, Maynooth University, Maynooth, Ireland;

\noindent{} 3. Department of Psychology, Maynooth University, Maynooth, Ireland;

\noindent{} $\ast$ Corresponding author; e-mail: gabriel.palma.2022@mumail.ie

\bigskip


\bigskip


\bigskip

\textit{Manuscript type}: Research paper. 

\bigskip

\noindent{\footnotesize Prepared using the suggested \LaTeX{} template for \textit{Am.\ Nat.}}

\newpage{}

\section*{Abstract}

Theta oscillations, ranging from 4-8 Hz, play a significant role in spatial learning and memory functions during navigation tasks. Frontal theta oscillations are thought to play an important role in spatial navigation and memory. Electroencephalography (EEG) datasets are very complex, making any changes in the neural signal related to behaviour difficult to interpret. However, multiple analytical methods are available to examine complex data structure, especially machine learning based techniques. These methods have shown high classification performance and the combination with feature engineering enhances the capability of these methods. This paper proposes using hidden Markov and linear mixed effects models to extract features from EEG data. Based on the engineered features obtained from frontal theta EEG data during a spatial navigation task in two key trials (first, last) and between two conditions (learner and non-learner), we analysed the performance of six machine learning methods (Polynomial Support Vector Machines, Non-linear Support Vector Machines, Random Forests, K-Nearest Neighbours, Ridge, and Deep Neural Networks) on classifying learner and non-learner participants. We also analysed how different standardisation methods used to pre-process the EEG data contribute to classification performance. We compared the classification performance of each trial with data gathered from the same subjects, including solely coordinate-based features, such as idle time and average speed. We found that more machine learning methods perform better classification using coordinate-based data. However, only deep neural networks achieved an area under the ROC curve higher than 80\% using the theta EEG data alone. Our findings suggest that standardising the theta EEG data and using deep neural networks enhances the classification of learner and non-learner subjects in a spatial learning task.

\newpage{}

\section{Introduction} 

Navigating from one place to the next is a complex cognitive skill that relies on the brain's ability to represent spatial information and retrieve it from memory. Studies in rodents and other animals have been instrumental in uncovering foundational mechanisms of spatial cognition and memory. The hippocampus, entorhinal cortex, and parietal cortex form the core of a widespread navigation circuit. Theta oscillations in the 4-8 Hz frequency range have been shown to play a critical role in spatial learning and memory during navigation tasks. Accumulating evidence has demonstrated the role of frontal midline theta in spatial learning and exploration \citep{roberts2013oscillatory, crespo2016slow, liang2021common, chrastil2022theta, du2023, thornberry2023} as well as successful retrieval \citep{klimesch1997theta, buzsaki2005theta, kaplan2012movement, roberts2013oscillatory, kaplan2014medial, greenberg2015decreases, lin2017theta, herweg2020theta}. It is possible that frontal theta oscillations facilitate communication between the hippocampus and the cortex to support the encoding of spatial memories \citep{buzsaki2005theta, mitchell2008frontal, buzsaki2013memory, kerren2018optimal, herweg2020theta, liang2021common}. 

However, analysing human scalp-EEG data collected during real-world or virtual spatial navigation poses challenges, due to the complexity and high dimensionality of the data. Machine learning techniques offer promising solutions by leveraging large datasets and automating the discovery of informative features. The Support Vector Machine (SVM) approach has proven useful in extracting features of theta oscillations involved in working memory retention \citep{johannesen2016machine}. The conformal kernel-based fuzzy support vector machine (CKF-SVM) has demonstrated high classification accuracy using frontal theta oscillations to differentiate between individuals with Mild Cognitive Impairment (MCI) \& healthy controls \citep{hsiao2021eeg}. Interestingly, event-related potentials (ERPs) elicited from a working memory auditory task were not predictive of cognitive performance. However, ERPs from a visual working memory task predicted information processing speed in Multiple Sclerosis patients and healthy controls \citep{kiiski2018machine}. Considering spatial navigation is highly visual, and oscillatory activity, as opposed to event-related potentials, shows greater promise in predictive ability \citep{vahid2018machine}, for this paper we have focussed primarily on theta (4-8 Hz) during a spatial learning and memory task. 

In this study, we aimed to develop an approach using hidden Markov models and mixed models to extract informative features from frontal midline theta EEG data collected during a virtual water maze task. We then evaluated multiple machine learning algorithms' ability to classify between learning and non-learner subjects based on the engineered theta features from early (encoding) and late (remembered) trials. Our goal was to determine a preprocessing and machine learning pipeline that can best decode neural signatures of spatial learning from EEG. We hope that this work will provide methodological advances and a more standardised, streamlined approach for analysing complex neural time series data without the need to evaluate various approaches. This work investigates the effectiveness of hidden Markov and linear mixed-effect models to extract features from theta EEG data. We hope to provide a standardised approach to predictive EEG analysis using spatial learning tasks to reduce time for neuroscientists and researchers in clinical settings.

\section{Methods}
\subsection{Experimental procedure}
Fifty adults (36 F, 14 M) aged between 18 and 45 ($\mbox{mean} = 21.7$) were recruited via the Maynooth University Department of Psychology and externally via social media and other methods. All participants gave informed consent before starting the experiment and were given a full briefing on the experiment and the exclusion criteria. Some participants from Maynooth University received course credit for participation. The experiment received ethical approval from the Maynooth University ethics committee. 
All participants undertook a computer-based spatial learning task which took place in a darkened, electrically-shielded and sound-attenuated testing cubicle (150 cm $\times$ 180 cm) with access to a joystick for navigating. The spatial navigation task used was NavWell (see \cite{commins2020navwell} for in-depth details), which consisted of a medium circular environment (15.75 seconds to traverse the arena, calculated at 75 Virtual Metres) through which participants could navigate. To aid navigation two cues were used and were located on the arena's wall: a yellow square (northeast quadrant wall) and a light of 50\% luminance. A square goal was hidden in the middle of the floor and was 15\% of the total arena size and consisted of a bright blue square that only became visible when the participant crossed it. 
Participants underwent 12 trials to try and find the hidden target. Participants were divided into two conditions, learner ($n = 25$) \& non-learner group ($n = 25$). The learner group had a maximum of 60 seconds per trial to find the hidden goal. There was a 10-second inter-trial interval between each trial to allow for rest. The non-learner group also had to navigate the arena but did not have a hidden goal. The non-learner group trials were time-matched to the average trial time of the learner group for accurate comparison and EEG signal processing. The X-Y coordinate data was recorded by the NavWell software from which distance, path length, idle time and other behavioural measures were extracted (see analysis below). Speed was kept constant across both conditions. The starting position for all trials was also kept constant across both conditions. For analysis, we only focussed on two trials (of the 12) for both groups – trial 1 (where neither group had learned the task) and trial 12 (where only the learner group should have learned the task).

\subsection{EEG Data Recording \& Extraction}
A BioSemi ActiveTwo system (BioSemi B.V., Amsterdam, Netherlands), which provided 32 Ag/AgCl electrodes, was positioned according to the 10/20 system, an international system denoting EEG electrode layout. This is the most common layout, meaning that the electrodes are either a distance of 10\% or 20\% from each other. Event triggers were sent for when participants began their trial and when they reached the goal or their trial ended. BioSemi-designed caps using the 32-electrode international 10-20 layout were also used. Eye movements and blinks were monitored using four external electrodes placed on the face. Raw EEG data were sampled at 1024 Hz but were down-sampled offline to 512 Hz.

The data were processed offline using the MATLAB-based software Brainstorm 12 \citep{tadel2011}. Data were pre-processed using a 1-40 Hz band-pass filter and were visually inspected for bad segments. Independent Component Analysis (ICA) was used to remove and/or correct artefacts in the data. EEG data were then referenced to the average of the 32 channels. Artefact-free data were then epoched for participants' full trial length, taking the entire time between the first two start/end events and the last two start/end events (cross-checked via the time reported in NavWell). These differed for the learner group but were standardised in the non-learner group due to the time-matching. We used a Morlet wavelet time-frequency analysis, with a central frequency of 1 Hz, a full-width half maximum time resolution of 3 seconds, and a linear frequency definition from 4 to 8 Hz (4:1:8). We then averaged across this frequency band and extracted the theta power at each time-point across the epoch for each participant at the frontal midline (averaged F3, Fz, and F4 electrodes) \cite{liang2021, thornberry2023, du2023}.

\subsection{Feature engineering}

The frontal theta waves dataset was composed of the total time the individual travelled in the experiment, the raw midline value of the theta wave, the subject ID, the group (learner or non-learner), and the trial indicator (trial 1 or trial 12). Moreover, the dataset containing the coordinates comprised the subject ID, the total time, $T$, the individual walked during the experiment, the $x$ coordinate, the $y$ coordinate at time $t$ (each coordinate was recorded every $0.25$ seconds), the group (learner or non-learner) and the trial (1 or 12).

Using the coordinates dataset, for each subject, we computed the \textit{total idle time} (the time that a subject did not move), \textit{total path length} (the journey's distance of the subjects), \textit{total angle shift} (the total angle changes for each subjects' step, calculated by the sum of absolute differences in angle shift, i.e. $$\displaystyle\sum_{t = 3}^{T}\left|\mbox{tan}^{-1}\left(\frac{y_t - y_{t-1}}{x_t -x_{t-1}}\right)-\mbox{tan}^{-1}\left(\frac{y_{t-1} - y_{t-2}}{x_{t-1} -x_{t-2}}\right)\right| \frac{180}{\pi},$$ where $\{x_t\}$ and $\{y_t\}$ are the time series of $x$ and $y$ coordinates for subject position), and \textit{average speed} (the total path length divided by the time to find the target). As an exploratory analysis, to identify differences among trials and groups, we first fitted Generalized Additive Models for Location, Scale and Shape (GAMLSS) \citep{Rigby2005, stasinopoulos2017} for each engineered feature. We modelled the location and scale parameters of a Gamma GAMLSS using the \textit{total angle shift}, \textit{average speed}, \textit{total idle time} and \textit{total angle shift} as predictors.


Let $x_t$ be the recorded theta power at time $t, t=1,\ldots,T$. We rescaled the theta power values using two types of standardisation. The first (\textit{minmax}) constrained the values to be between 0 and 1 through  $$x^{\text{minmax}}_t = \frac{x_t - \min(\mathbf{x})}{\max(\mathbf{x}) - \min(\mathbf{x})}.$$ The second involves a \textit{Z-score} transformation, such that $$x^{\text{Z-score}}_t=\frac{x_t-\bar{x}}{s},$$ where $\bar{x}$ is the mean and $s$ is the standard deviation of the sample.

For the $x_t$, $x_t^{\text{minmax}}$, and $x_t^{\text{Z-score}}$ data, we extracted and engineered a set of different features. We computed the height and curvature (Calculated by taking the second-order difference of $x_{k-1}, x_{k}, x_{k+1}$, where $k$ is the time where a peak occurred) of each peak within the EEGs for each subject. After that, we fitted a linear mixed-effects model (LMM) to the peak heights, including random intercepts and slopes over peak curvature per participant. We then extracted the predicted random intercepts and slopes (one intercept and slope per participant) and used them as features in the machine learning methods described in the later sections.


In addition to that, to extract additional features from the EEG theta signals for each participant, we fitted Gaussian Hidden Markov models (HMMs) \cite{zucchini2016} to each EEG. HMMs can be used to model time series data assuming there are latent states which determine the mean and variance of the time series at different stages. Let $C_t\in\{S_1,S_2,\ldots,S_M\}$ be a categorical variable with $M$ categories, describing the latent state of the series at time $t$. We assume the Markov property of order 1, which means that the state of the series at $t-1$ influences the state at time $t$. In algebraic notation, we have $$\mbox{P}(C_t=c_t|C_{t-1},C_{t-2},\ldots,C_1)=\mbox{P}(C_t=c_t|C_{t-1}).$$ We then formulate a HMM with $M=4$ states to be used to analyse the EEG data, by assuming that $X_t|X_{t-1},C_t \sim \mbox{N}(\mu(C_t),\sigma^2(C_t))$.
We estimate the mean and variance for each latent state, and they are used to determine the mean $\mu_t=\mu(C_t)$ and variance $\sigma^2_t=\sigma^2(C_t)$ of the EEG time series process at time $t$. One important feature of HMMs is the transition probability matrix $\mathbf{P}$ that is estimated from the data. This matrix governs the likelihood of switching from one state to another, or remaining in the same state, given the state at the previous time point. Since we are assuming 4 states, we have
\begin{eqnarray*}
    \mathbf{P}&=&\left(\begin{array}{cccc}
        \pi_{11} & \pi_{12} & \pi_{13} & \pi_{14} \\
        \pi_{21} & \pi_{22} & \pi_{23} & \pi_{24} \\
        \pi_{31} & \pi_{32} & \pi_{33} & \pi_{34} \\
        \pi_{41} & \pi_{42} & \pi_{43} & \pi_{44}
    \end{array}\right),
\end{eqnarray*}
where $\pi_{ij}$ is the probability of the series switching from state $i$ to state $j$, $i,j\in\{1,2,3,4\}$.





For each subject presented in the study, we estimated the means and variances for all four states, as well as the transition probabilities. This totals us eight parameter estimates (four means and four variances) per participant. In addition, we calculated how frequent each state was in the series for each participant, adding three extra features for states 1, 2 and 3 (since the frequency for state four is one minus the frequencies for states 1, 2 and 3). The choice of four states was made based on previous exploration of model fits through the Akaike information criterion (AIC); we present the results for other values of $M$ as Supplementary Material.

Estimation was done using the EM algorithm implemented through package depmixS4~\cite{Visser2010} available for R software~\cite{R2022}.

\subsection*{Learner and non-learner classification}
 
We created two primary datasets to train different machine learning methods to classify the EEG time series as arising either from a participant in the non-learner or learner group. The \textbf{EEG data} contains the time series features, HMM and LMM parameter estimates. The \textbf{coordinates} data solely contains variables obtained from the coordinates dataset. 

To identify the effect of the selected features on the classification performance, we used 3rd order Polynomial Support Vector Machines (Poly SVM), Non-linear Support Vector Machines (Non-linear SVM), Random Forests (RF) with one thousand trees and a depth of 5, K-Nearest Neighbours (KNN) with one neighbour, elastic net regularisation in logistic regression with $\alpha = 0.98$ (constant that multiplies the L2 regularisation), and Deep Neural Networks (DNN) with eight layers containing $100, 150, 200, 150, 46, 20, 10$ and one neuron per layer. We evaluated the performance of each machine learning algorithm using Leave-One-Out Cross-Validation (LOOCV).

After selecting the best model trained with the EEG dataset, we used the Local Interpretable Model-agnostic Explanations (LIME) algorithm to extract feature importance. To visualise the feature importance for each prediction of the best learning algorithm within the step of LOOCV, we obtain the feature importance for every prediction related to a subject in our dataset. Finally, with this list of features' importance per subject, we list the top three most frequent ones for both trials and groups.

\section{Results}
In Section~\ref{EngineeredFeaturesAnalysis}, we present the results of the analysis of the engineered features based on coordinates data and the EEG data. We also present the overall performance of all machine learning algorithms for each number of states, $M$, of the hidden Markov Model. In Section~\ref{ClassificationResults}, we present the detailed performance of the machine learning algorithms for classifying non-learner and learner subjects using $M=4$.

\subsection{Analysis of Engineered Features\label{EngineeredFeaturesAnalysis}}

\begin{figure}[!ht]
    \centering
    \includegraphics[width=1\textwidth]{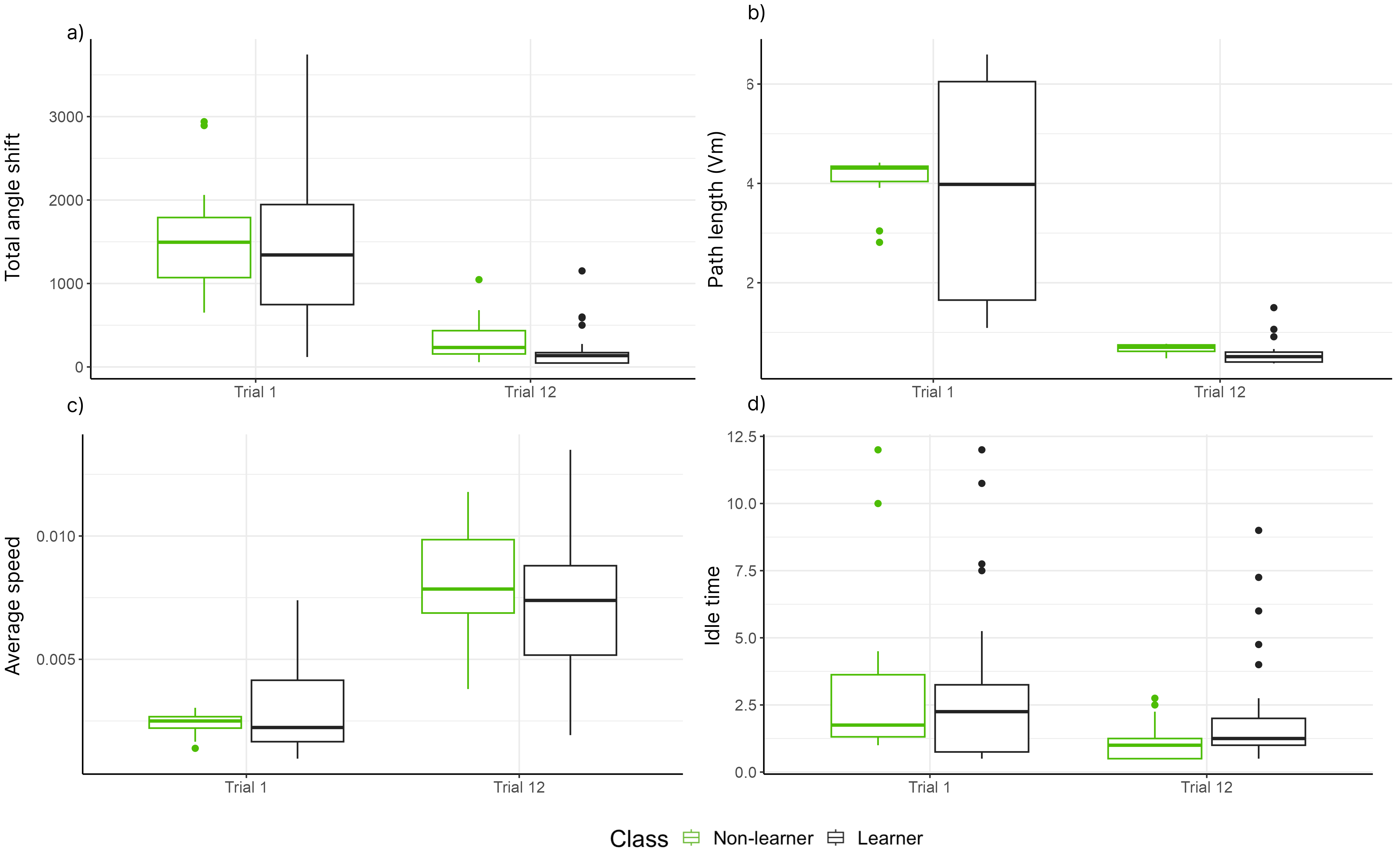}
    \caption{Behavioural findings based on the coordinate data: a) Total angle shift; b) Path length (Vm); c) Average speed; d) Idle time. The coordinates of all participants are presented from trials one and twelve, and the colour indicates the groups of the groups.}
    \label{FeatureEngeneeredPlots}
\end{figure}

Figure~\ref{FeatureEngeneeredPlots} illustrates the effect of trials and groups on the engineered features based on the coordinates data. After fitting the Generalized Additive Models for Location, Scale and Shape for each feature, our results showed that, for all engineered features, modelling the mean and dispersion of the Gamma distribution as a function of trial and group is best, based on AIC. For the total angle shift (Figure~\ref{FeatureEngeneeredPlots}a), a significant difference between trials (LR = $72.32$, df = $1$, $p < 0.01$), no differences between groups (LR = $2.19$, df = $1$, $p = 0.13$) and no interaction between trial and groups (LR = $1.49$, df = $1$, $p = 0.22$) were found. 

For the path length (Figure~\ref{FeatureEngeneeredPlots}b), no interaction was found for the mean of the Gamma distribution (LR = $0.46$, df = $1$, $p = 0.49$). Also, differences between trials (LR = $193.42$, df = $1$, $p < 0.01$) and groups (LR = $4.74$, df = $1$, $p = 0.029$) were found. For the average speed (Figure~\ref{FeatureEngeneeredPlots}c), we found differences between trials (LR = $102.83$, df = $1$, $p < 0.01$), no difference between groups (LR = $1.35$, df = $1$, $p = 0.24$) and no interaction between groups and trials LR = $3.66$, df = $1$, $p = 0.056$). Finally, for idle time (Figure~\ref{FeatureEngeneeredPlots}d), there is an interaction between trials and groups for the mean parameter of the Gamma distribution (LR = $4.09$, df = $1$, $p = 0.043$). There is difference between trials (LR = $22.47$, df = $2$, $p < 0.01$) and groups (LR = $10.13$, df = $2$, $p < 0.01$). For subjects in trial 1, there is no difference between groups (LR = $0.011$, df = $1$, $p = 0.91$) and for trial 12, there is a difference between groups (LR = $10.21$, df = $1$, $p = 0.001$). For non-learners, there is a difference between trials (LR = $20.49$, df = $1$, $p < 0.01$), and for learners, there is no difference between trials (LR = $2.16$, df = $1$, $p=0.14$).

\begin{figure}[!ht]
    \centering
    \includegraphics[width=1\textwidth]{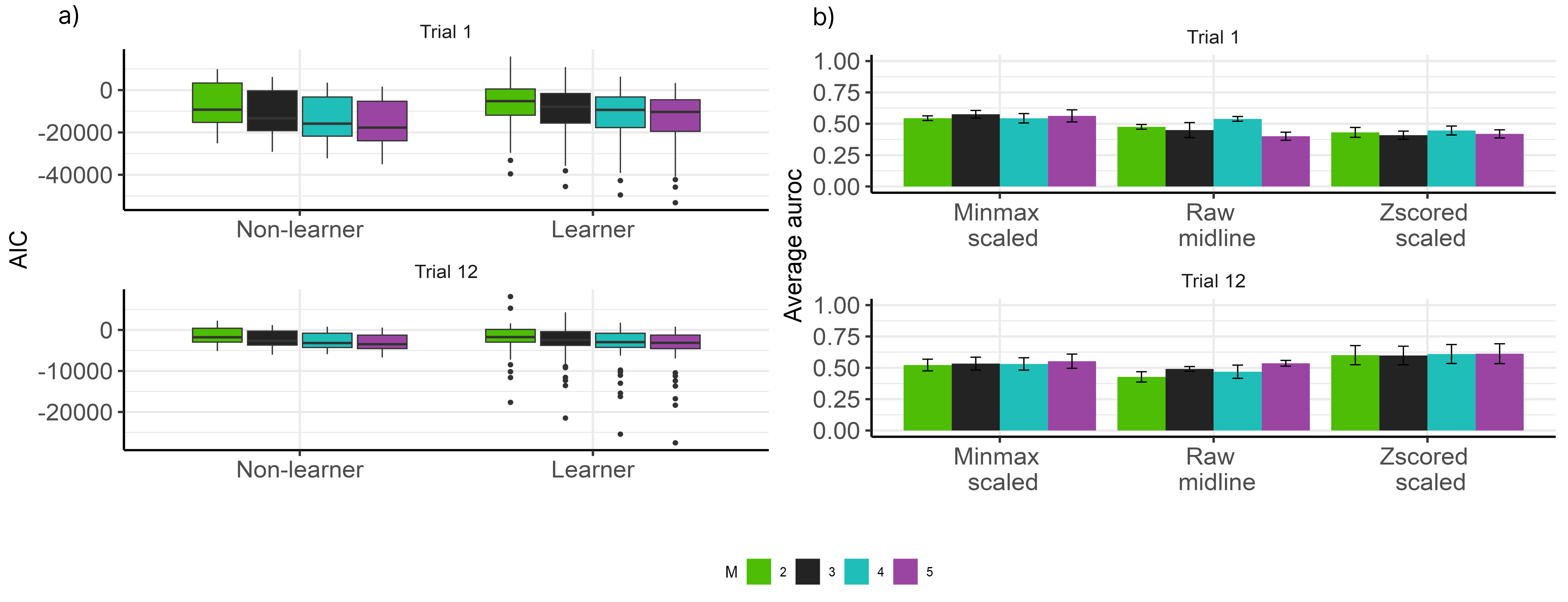}
    \caption{a) Box plots of the computed Akaike information criterion (AIC) from the hidden Markov models using $M = 2, 3, 4, 5$. Each point of the plot represents an AIC value for a hidden Markov model fitted using the EEG data of a subject. b) Average AUROC for all machine learning algorithms using the EEG data for each value of $M$.}
    \label{HMMPerformanceAIC}
\end{figure}



We then fitted the hidden Markov models for each participant for the respective groups and trials using EEG data. Figure~\ref{HMMPerformanceAIC}a shows the performance of the Gaussian hidden Markov model based on AIC. It illustrates no clear difference among the different values of $M$. Also, Figure~\ref{HMMPerformanceAIC}b shows that the average AUROC of the machine learning algorithm using different features based on the number of states $M$ also showed no clear difference. This finding supports the decision to solely present the performance of the selected learning algorithms with $M = 4$. The additional plots and code for reproducing them are available at \url{https://github.com/GabrielRPalma/UnderstandingLearningWithML}.

\begin{figure}[!ht]
    \centering
    \includegraphics[width=14cm]{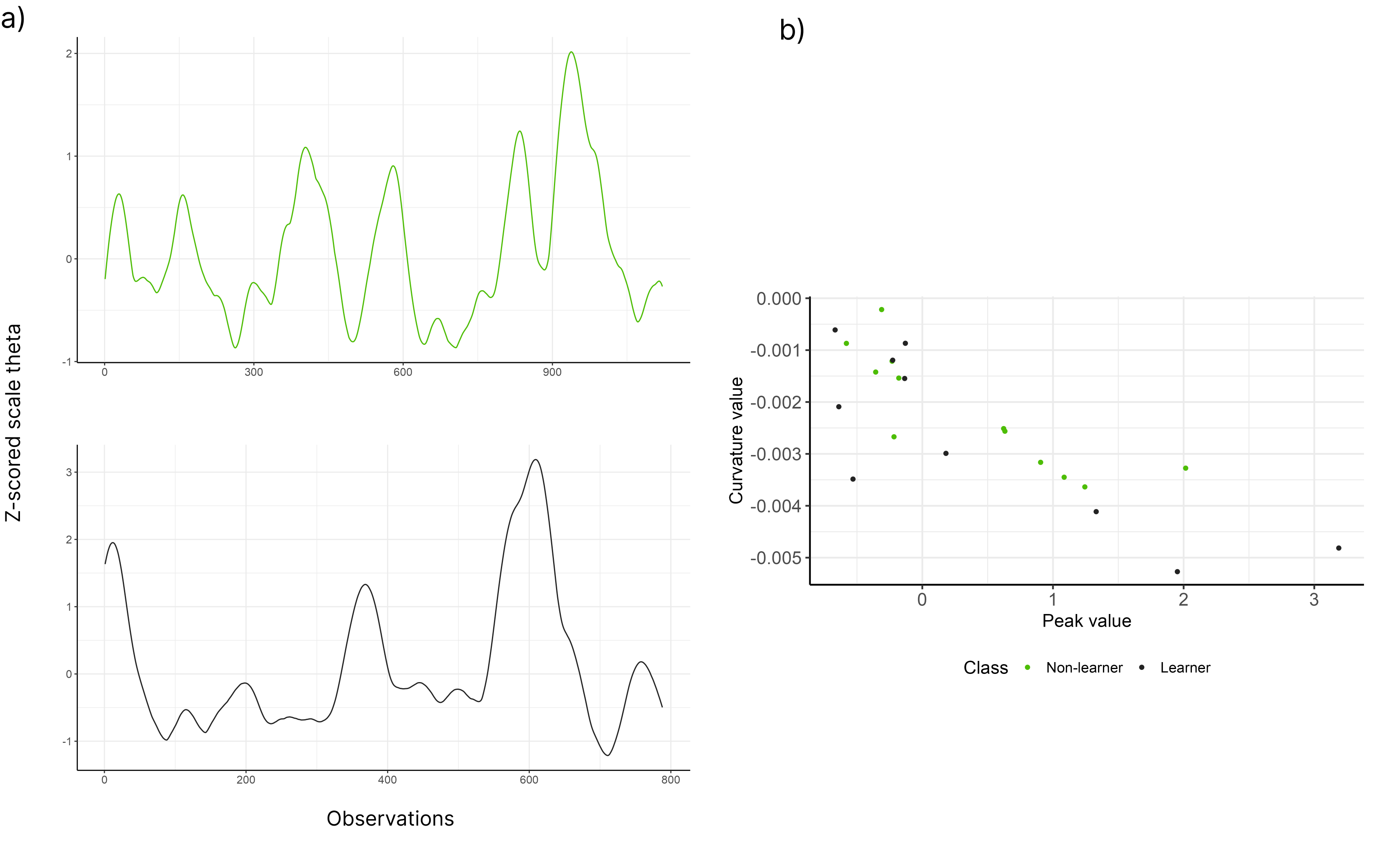}
    \caption{a) Z-score scaled theta time series from trial 12 for a non-learner (top) and learner (bottom). b) A scatter plot of the peak and curvature values extracted from each theta wave time series for the two subjects in a).}
    \label{EEG_features_Trial1}
\end{figure}

Finally, the association between the peak and curvature of the peak obtained from the Z-score scaled theta time series was obtained using a linear mixed-effects model. The features used to fit the linear mixed-effect model are illustrated in Figure~\ref{EEG_features_Trial1}. The slope and intercept of the model will be used as features for the machine learning classification of non-learner and learner (learner) groups in section~\ref{ClassificationResults}.

\subsection{Classifying Learning \label{ClassificationResults}}

\begin{figure}[!ht]
    \centering
    \includegraphics[width=1\textwidth]{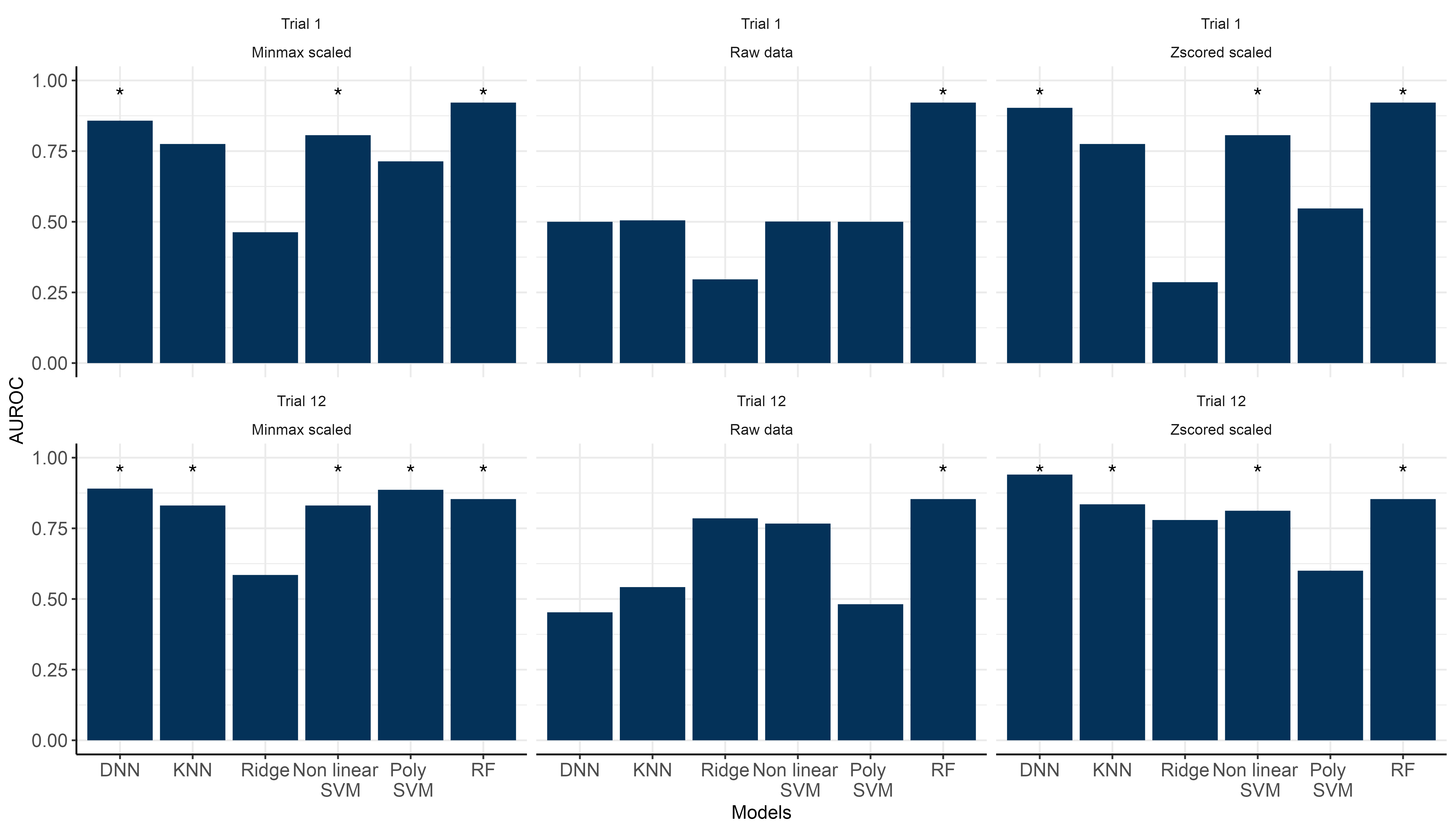}
    \caption{Area under the ROC (AUROC) curve obtained using all machine learning algorithms when classifying non-learner and learner subjects for Trial 1 and 12 solely using the coordinates data. \textbf{*} represents methods that achieved $\mbox{AUROC} > 0.8$.}
    \label{CoordinatesMLPerformance}
\end{figure}

Figure~\ref{CoordinatesMLPerformance} show the performance of the selected learning algorithms to classify non-learner or learner participants for both trials based solely on combined coordinate data. Here, we find that most machine learning algorithms perform well (with the exception of Ridge) at classifying whether a participant is a learner or non-learner. Random forests (RF), deep neural networks (DNN) and non-linear SVM perform particularly well (all with an AUROC larger than 0.8). Furthermore, the algorithms were better at classifying participants on Trial 12 compared to Trial 1. Finally, pre-processing the coordinates data using the Z-score and minimum and maximum standardisation improved most algorithms' performances, especially when compared to the raw data. 

\begin{figure}[!ht]
    \centering
    \includegraphics[width=1\textwidth]{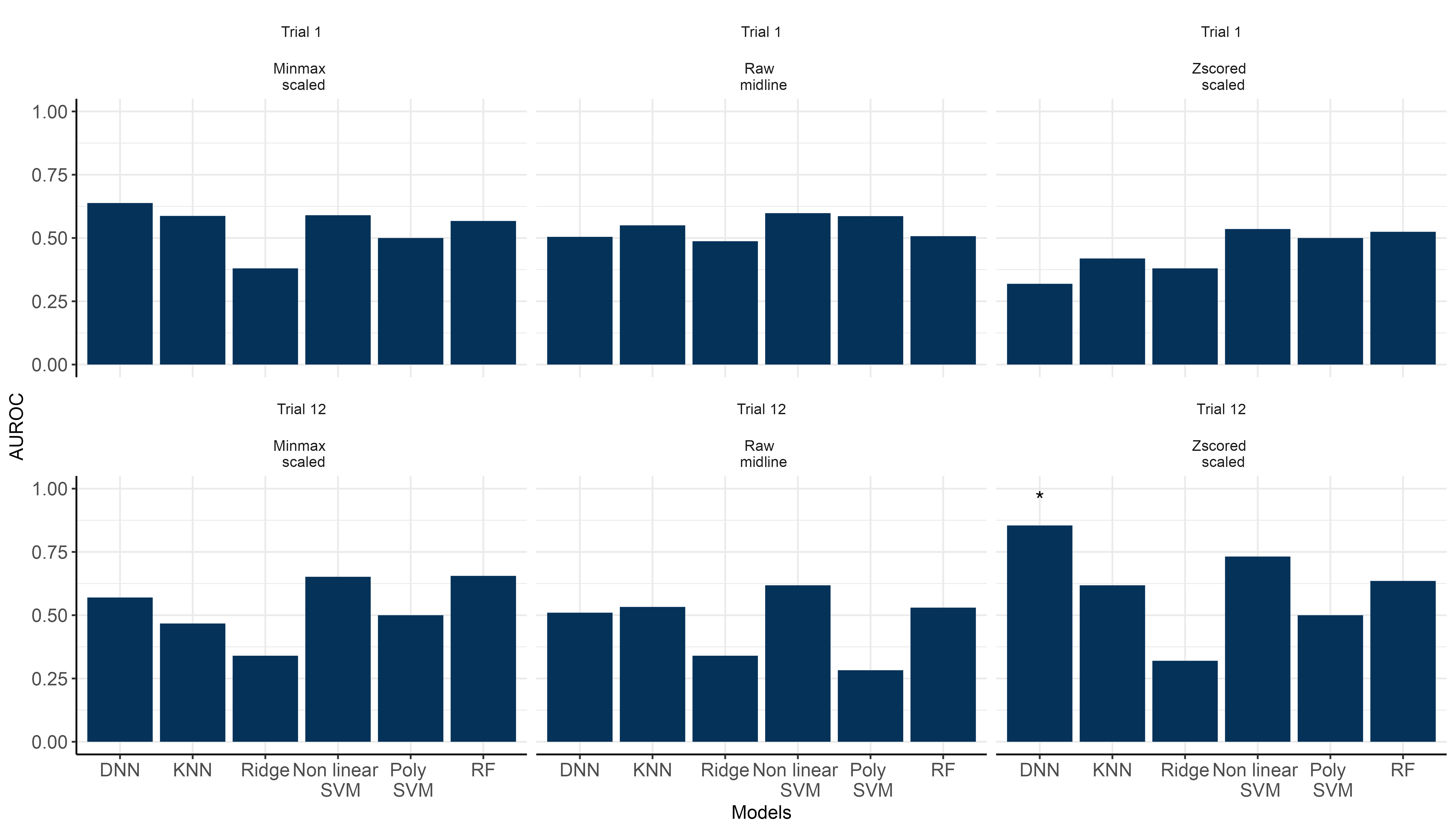}
    \caption{Area under the ROC (AUROC) curve obtained using all machine learning algorithms when classifying non-learner and learner subjects for Trial 1 and 12 solely using the EEG data. \textbf{*} represents methods that achieved $\mbox{AUROC} > 0.8$.}
    \label{EEGDataMLPerformance}
\end{figure}

Figure~\ref{EEGDataMLPerformance} shows the performance of the machine learning algorithms using the EEG dataset. Compared to using the coordinates data, the algorithms perform much worse. On Trial 1, most ML algorithms achieve AUROCs lower than 0.5 irrespective of the dataset used. While there is a general improvement across all algorithms on Trial 12, only the DNN achieved an AUROC larger than 0.8. This is noted particularly when using Z-score scaling to pre-process the data. 

\begin{figure}[!ht]
    \centering
    \includegraphics[width=1\textwidth]{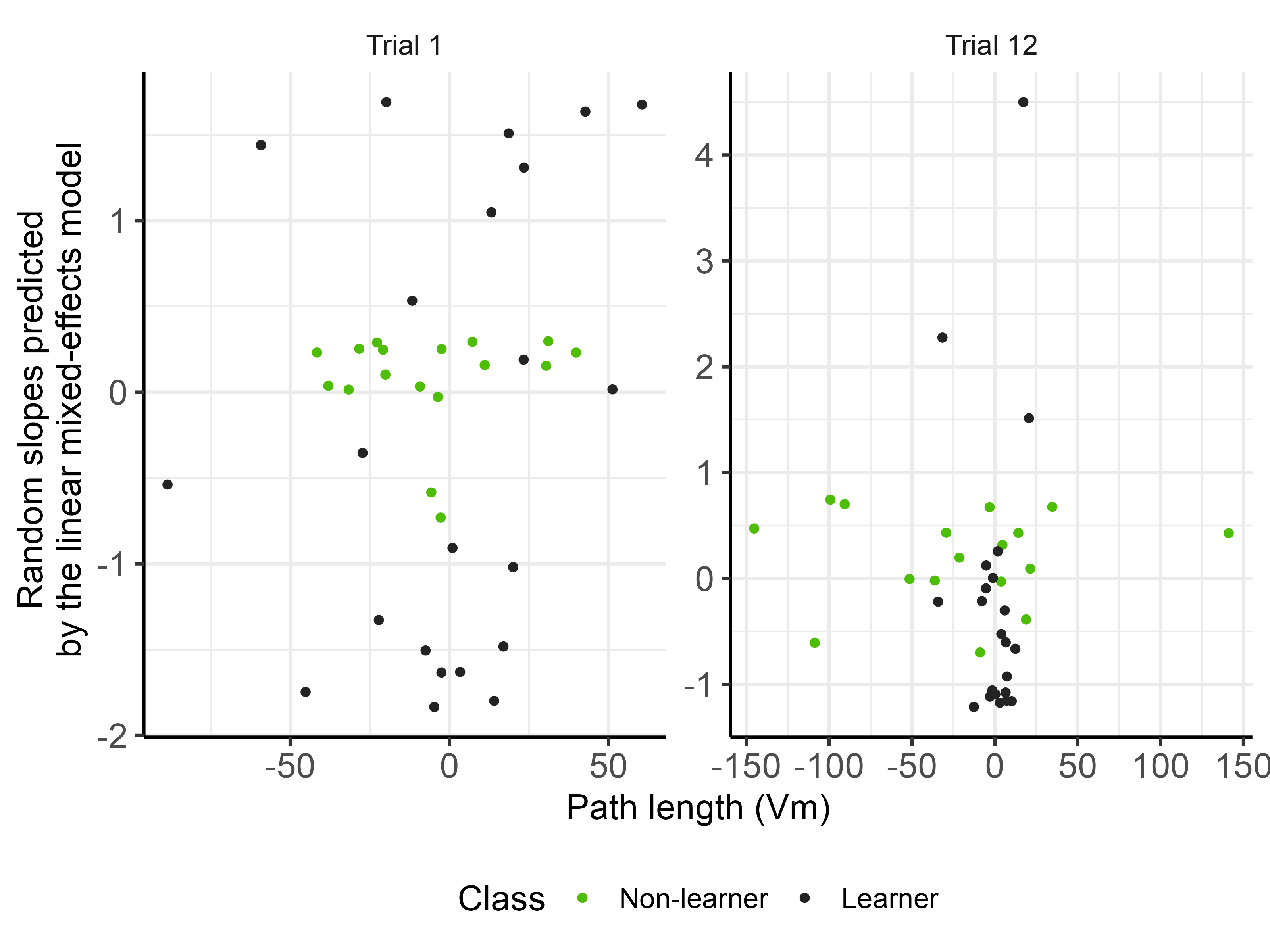}
    \caption{Scatter plot of the linear mixed-effect model slope and path length (Vm) for all subjects of trials 1 and 12. The slope was obtained based on the Z-score scaled EEG data. The colours represent the groups of each subject.}
    \label{LIME_feature_plot}
\end{figure}

The findings that DNN can discriminate between learners and non-learners on Trial 12 suggests that there might be something within the EEG pattern that can help distinguish between the two groups. To this end, we used the Local Interpretable Model-agnostic Explanations (LIME) method in an attempt to determine the key features (coordinate and EEG) that may help with the classification for both Trial 1 and Trial 12. Table~\ref{LIME_coefficients} presents the top 3 most frequent features with the relative weights selected for both groups and the two trials. On Trial 1, both EEG and coordinate features are ranked highly, specifically the random slopes from the linear mixed-effects model and total distance, respectively. By Trial 12, only the random slopes of the EEG data are ranked in the top 3. This feature emerges for both the learner and non-learner groups. Figure~\ref{LIME_feature_plot} shows a scatter plot of the features the LIME algorithm indicates. 

\begin{table}[]
    \centering
    \begin{tabular}{cccc}
        \hline
         Group & Trial & Feature & LIME coefficient\\
         \hline
          &   &  $\mbox{Linear mixed-effect model's slope} $ & 0.52\\[-0.1cm]
          Non-learner& Trial 1  &   $\mbox{Path length (Vm)}$ & 0.38\\[-0.1cm]
          &   &   $\mbox{Path length (Vm)}$ & 0.34\\[0.2cm]

          &   &  $\mbox{Linear mixed-effect model's slope}$ & 0.53\\[-0.1cm]
          Learner& Trial 1  &   $\mbox{Linear mixed-effect model's slope}$ & 0.52\\[-0.1cm]
          &   &   $\mbox{Path length (Vm)}$ & 0.46\\[0.3cm]

          &   &  $\mbox{Linear mixed-effect model's slope}$ & 0.75\\[-0.1cm]
          Non-learner & Trial 12  &   $\mbox{Linear mixed-effect model's slope}$ & 0.45\\[-0.1cm]
          &   &   $\mbox{Linear mixed-effect model's slope}$ & 0.40\\[0.2cm]

          &   &  $\mbox{Linear mixed-effect model's slope}$ & 0.58\\[-0.1cm]
          Learner & Trial 12  &   $\mbox{Linear mixed-effect model's slope} $ & 0.50\\[-0.1cm]
          &   &   $ \mbox{Linear mixed-effect model's slope}$ & 0.40\\
          \hline
    \end{tabular}
    \caption{Top 3 features with higher weights based on Local Interpretable Model-agnostic Explanations (LIME) method algorithm for the decision of the deep neural network trained with the Z-score scaled EEG data combined with the coordinate data.}
    \label{LIME_coefficients}
\end{table}

\section{Discussion}
In this paper, we proposed using hidden Markov and linear mixed-effect models to extract features from EEG theta time series. Our analysis showed promising results of deep neural networks for classifying non-learner and learner groups based on the engineered features collected from the EEG data. This finding points towards using deep learning-based methods for classifying spatial learning and memory processes based on theta time series. In addition, our findings indicate that the pre-processing method influences the learning algorithm selection for this task. Therefore, the Z-score transformation combined with deep neural networks allows for better performance when compared to the other machine learning methods. Other papers have demonstrated the effectiveness of deep learning algorithms for classification tasks based on EEG data \citep{ nirabi2021, tang2022}, which agree with the findings reported here. 

Based on our experimental paradigm, we would expect machine learning algorithms to perform worse classifying subjects on Trial 1 compared to Trial 12. In Trial 1, both groups of subjects are randomly searching for the target, since they have no prior experience with the task. However, by Trial 12 the learner group have learned and can successfully recall the target location, whereas the non-learner group have had no exposure to a target and are still randomly searching. 

The deep neural network was the only machine learning approach that could accurately demonstrate this expected pattern of poor Trial 1 classification performance (insert stats) with accurate Trial 12 classification performance (insert stats). Other models were not accurate enough to capture the underlying neural changes reflecting learning using EEG data in isolation, with most still not improving with the addition of coordinate data.

Therefore, the ability of the DNN to accurately match the expected neural changes demonstrates the potential of deep learning methods. As learners transition from random searching to spatial memory-guided navigation, deep neural networks appear to detect associated EEG changes. Recent studies have shown deep learning models can predict learning-related performance across trials using EEG data \citep{kang2020eeg, zygierewicz2022decoding}. Our findings fit with this literature, suggesting the potential of deep neural networks and hidden Markov models to decode spatial learning and memory processes.

In regards to model interpretation, the Local Interpretable Model-agnostic Explanations (LIME) method was selected to obtain a local fidelity interpretation for the decision made by the deep neural networks algorithm, given its best performance for classifying non-learner and learner using solely the proposed features based on the EEG dataset. Given that the LIME method provides a local regression based on K-Lasso, we presented the coefficients with higher weight provided by the method and the respective features used for classifying a subject.

Other researchers have reported variable importance based on LIME \cite{ribeiro2016should}, and it was well received by the machine learning community. Other explainable artificial intelligence (XAI) methods are constantly being developed, given the active research community built around this area \cite{longo2023explainable}. However, our goal in this paper was to provide a list of possible important variables used for a decision made by a deep neural network algorithm, and LIME was suitable for such a task.

Finally, our findings would support the theory that frontal midline theta power is involved in spatial learning and memory processes. For example, \cite{du2023frontal} recently reported that frontal-midline theta is involved in the early encoding of spatial information during active navigation (also see \cite{chrastil2022}). In addition to \cite{du2023frontal}, we also report that there is enough information contained within frontal midline theta during active spatial learning and subsequent memory-based navigation to facilitate accurate classification of learner and non-learner subjects. Importantly, frontal midline theta may provide a non-invasive detection method for spatial memory or cognition difficulties. This would be incredibly useful as an early detector of spatial impairment for those with pre-clinical Alzheimer’s disease, as this symptom is often reported early before formal diagnosis~\citep{kunz2015, coughlan2018, coughlan2020}. Relative theta power at rest has been used to discriminate between Alzheimer's disease patients and healthy controls~\citep{musaeus2018}. However, including a greater age demographic and analysis of other regions known to contribute to spatial memory using our proposed technique would be required to validate our findings. Additionally, task-related or goal-directed FM-theta may only be useful in predicting spatial learning. The method proposed in this paper should be applied to other tasks and experimental paradigms to support our findings further.

\section{Conclusion}
A new approach was proposed to extract features from EEG theta time series based on linear mixed-effects and hidden Markov models. We showed that the z-score type transformation of EEG theta time series combined with the flexibility of deep neural networks can achieve better performance for classifying non-learner and learner individuals. Therefore, recommendations on feature engineering of EEG data and pre-processing approaches on EEG based on theta time series can be given to researchers who aim to classify the learning stages using a machine learning approach. This work forms a basis for further studies interested in investigating learning effects based on EEG theta time series. 

\section{Acknowledgments}

This publication has emanated from research conducted with the financial support of Funda\c{c}\~{a}o de Amparo \`{a} Pesquisa do Estado de S\~{a}o Paulo (proc. no. 19/14805-7 and no. 20/06147-7), Ag\^{e}ncia USP de Inova\c{c}\~{a}o and Science Foundation Ireland under Grant 18/CRT/6049. The opinions, findings and conclusions or recommendations expressed in this material are those of the authors and do not necessarily reflect the views of the funding agencies.

\section{Declarations}
~~~~
\textbf{Ethical Approval} The use of human subjects with EEG was approved by the Maynooth University Biomedical \& Life Sciences Research Ethics Subcommittee (BSRESC-2021-2453422).

\textbf{Competing interests} Not applicable.

\textbf{Authors’ contributions} All authors conceived and designed the research. C.T. collected the data and provided insights into the discussion of results. G.R.P. and R.A.M. created the feature engineering methodology and analysed the data. G.R.P. led the writing of the manuscript. All authors contributed to the overall writing. 

\textbf{Funding} This publication has emanated from research conducted with the financial support of Science Foundation Ireland under Grant number 18/CRT/6049.

\textbf{Availability of data and materials} All datasets and scripts are made available at \url{https://github.com/GabrielRPalma/UnderstandingLearningWithML}

\bibliographystyle{apalike}
\bibliography{ref}

\begin{thebibliography}{}

\bibitem[Buzs{\'a}ki, 2005]{buzsaki2005theta}
Buzs{\'a}ki, G. (2005).
\newblock Theta rhythm of navigation: link between path integration and
  landmark navigation, episodic and semantic memory.
\newblock {\em Hippocampus}, 15(7):827--840.

\bibitem[Buzs{\'a}ki and Moser, 2013]{buzsaki2013memory}
Buzs{\'a}ki, G. and Moser, E.~I. (2013).
\newblock Memory, navigation and theta rhythm in the hippocampal-entorhinal
  system.
\newblock {\em Nature neuroscience}, 16(2):130--138.

\bibitem[Chrastil et~al., 2022a]{chrastil2022theta}
Chrastil, E.~R., Rice, C., Goncalves, M., Moore, K.~N., Wynn, S.~C., Stern,
  C.~E., and Nyhus, E. (2022a).
\newblock Theta oscillations support active exploration in human spatial
  navigation.
\newblock {\em NeuroImage}, 262:119581.

\bibitem[Chrastil et~al., 2022b]{chrastil2022}
Chrastil, E.~R., Rice, C., Goncalves, M., Moore, K.~N., Wynn, S.~C., Stern,
  C.~E., and Nyhus, E. (2022b).
\newblock Theta oscillations support active exploration in human spatial
  navigation.
\newblock {\em NeuroImage}, 262:119581.

\bibitem[Commins et~al., 2020]{commins2020navwell}
Commins, S., Duffin, J., Chaves, K., Leahy, D., Corcoran, K., Caffrey, M.,
  Keenan, L., Finan, D., and Thornberry, C. (2020).
\newblock Navwell: A simplified virtual-reality platform for spatial navigation
  and memory experiments.
\newblock {\em Behavior research methods}, 52:1189--1207.

\bibitem[Coughlan et~al., 2018]{coughlan2018}
Coughlan, G., Lacz{\'o}, J., Hort, J., Minihane, A.-M., and Hornberger, M.
  (2018).
\newblock Spatial navigation deficits—overlooked cognitive marker for
  preclinical alzheimer disease?
\newblock {\em Nature Reviews Neurology}, 14(8):496--506.

\bibitem[Coughlan et~al., 2020]{coughlan2020}
Coughlan, G., Puthusseryppady, V., Lowry, E., Gillings, R., Spiers, H.,
  Minihane, A.-M., and Hornberger, M. (2020).
\newblock Test-retest reliability of spatial navigation in adults at-risk of
  alzheimer’s disease.
\newblock {\em PLoS One}, 15(9):e0239077.

\bibitem[Crespo-Garc{\'\i}a et~al., 2016]{crespo2016slow}
Crespo-Garc{\'\i}a, M., Zeiller, M., Leupold, C., Kreiselmeyer, G., Rampp, S.,
  Hamer, H.~M., and Dalal, S.~S. (2016).
\newblock Slow-theta power decreases during item-place encoding predict spatial
  accuracy of subsequent context recall.
\newblock {\em Neuroimage}, 142:533--543.

\bibitem[Du et~al., 2023a]{du2023}
Du, Y.~K., Liang, M., McAvan, A.~S., Wilson, R.~C., and Ekstrom, A.~D. (2023a).
\newblock Frontal-midline oscillations index the evolution of spatial memory
  during active navigation.
\newblock {\em bioRxiv}, pages 2023--04.

\bibitem[Du et~al., 2023b]{du2023frontal}
Du, Y.~K., Liang, M., McAvan, A.~S., Wilson, R.~C., and Ekstrom, A.~D. (2023b).
\newblock Frontal-midline theta and posterior alpha oscillations index early
  processing of spatial representations during active navigation.
\newblock {\em Cortex}, 169:65--80.

\bibitem[Greenberg et~al., 2015]{greenberg2015decreases}
Greenberg, J.~A., Burke, J.~F., Haque, R., Kahana, M.~J., and Zaghloul, K.~A.
  (2015).
\newblock Decreases in theta and increases in high frequency activity underlie
  associative memory encoding.
\newblock {\em Neuroimage}, 114:257--263.

\bibitem[Herweg et~al., 2020]{herweg2020theta}
Herweg, N.~A., Solomon, E.~A., and Kahana, M.~J. (2020).
\newblock Theta oscillations in human memory.
\newblock {\em Trends in cognitive sciences}, 24(3):208--227.

\bibitem[Hsiao et~al., 2021]{hsiao2021eeg}
Hsiao, Y.-T., Wu, C.-T., Tsai, C.-F., Liu, Y.-H., Trinh, T.-T., and Lee, C.-Y.
  (2021).
\newblock Eeg-based classification between individuals with mild cognitive
  impairment and healthy controls using conformal kernel-based fuzzy support
  vector machine.
\newblock {\em International Journal of Fuzzy Systems}, 23:2432--2448.

\bibitem[Johannesen et~al., 2016]{johannesen2016machine}
Johannesen, J.~K., Bi, J., Jiang, R., Kenney, J.~G., and Chen, C.-M.~A. (2016).
\newblock Machine learning identification of eeg features predicting working
  memory performance in schizophrenia and healthy adults.
\newblock {\em Neuropsychiatric electrophysiology}, 2:1--21.

\bibitem[Kang et~al., 2020]{kang2020eeg}
Kang, T., Chen, Y., Fazli, S., and Wallraven, C. (2020).
\newblock Eeg-based prediction of successful memory formation during vocabulary
  learning.
\newblock {\em IEEE Transactions on Neural Systems and Rehabilitation
  Engineering}, 28(11):2377--2389.

\bibitem[Kaplan et~al., 2014]{kaplan2014medial}
Kaplan, R., Bush, D., Bonnefond, M., Bandettini, P.~A., Barnes, G.~R., Doeller,
  C.~F., and Burgess, N. (2014).
\newblock Medial prefrontal theta phase coupling during spatial memory
  retrieval.
\newblock {\em Hippocampus}, 24(6):656--665.

\bibitem[Kaplan et~al., 2012]{kaplan2012movement}
Kaplan, R., Doeller, C.~F., Barnes, G.~R., Litvak, V., D{\"u}zel, E.,
  Bandettini, P.~A., and Burgess, N. (2012).
\newblock Movement-related theta rhythm in humans: coordinating self-directed
  hippocampal learning.
\newblock {\em PLoS biology}, 10(2):e1001267.

\bibitem[Kerr{\'e}n et~al., 2018]{kerren2018optimal}
Kerr{\'e}n, C., Linde-Domingo, J., Hanslmayr, S., and Wimber, M. (2018).
\newblock An optimal oscillatory phase for pattern reactivation during memory
  retrieval.
\newblock {\em Current Biology}, 28(21):3383--3392.

\bibitem[Kiiski et~al., 2018]{kiiski2018machine}
Kiiski, H., Jollans, L., Donnchadha, S.~{\'O}., Nolan, H., Lonergan, R., Kelly,
  S., O’Brien, M.~C., Kinsella, K., Bramham, J., Burke, T., et~al. (2018).
\newblock Machine learning eeg to predict cognitive functioning and processing
  speed over a 2-year period in multiple sclerosis patients and controls.
\newblock {\em Brain topography}, 31:346--363.

\bibitem[Klimesch et~al., 1997]{klimesch1997theta}
Klimesch, W., Doppelmayr, M., Schimke, H., and Ripper, B. (1997).
\newblock Theta synchronization and alpha desynchronization in a memory task.
\newblock {\em Psychophysiology}, 34(2):169--176.

\bibitem[Kunz et~al., 2015]{kunz2015}
Kunz, L., Schr{\"o}der, T.~N., Lee, H., Montag, C., Lachmann, B., Sariyska, R.,
  Reuter, M., Stirnberg, R., St{\"o}cker, T., Messing-Floeter, P.~C., et~al.
  (2015).
\newblock Reduced grid-cell--like representations in adults at genetic risk for
  alzheimer’s disease.
\newblock {\em Science}, 350(6259):430--433.

\bibitem[Liang et~al., 2021a]{liang2021common}
Liang, M., Zheng, J., Isham, E., and Ekstrom, A. (2021a).
\newblock Common and distinct roles of frontal midline theta and occipital
  alpha oscillations in coding temporal intervals and spatial distances.
\newblock {\em Journal of Cognitive Neuroscience}, 33(11):2311--2327.

\bibitem[Liang et~al., 2021b]{liang2021}
Liang, M., Zheng, J., Isham, E., and Ekstrom, A. (2021b).
\newblock Common and distinct roles of frontal midline theta and occipital
  alpha oscillations in coding temporal intervals and spatial distances.
\newblock {\em Journal of Cognitive Neuroscience}, 33(11):2311--2327.

\bibitem[Lin et~al., 2017]{lin2017theta}
Lin, J.-J., Rugg, M.~D., Das, S., Stein, J., Rizzuto, D.~S., Kahana, M.~J., and
  Lega, B.~C. (2017).
\newblock Theta band power increases in the posterior hippocampus predict
  successful episodic memory encoding in humans.
\newblock {\em Hippocampus}, 27(10):1040--1053.

\bibitem[Longo, 2023]{longo2023explainable}
Longo, L. (2023).
\newblock {\em Explainable Artificial Intelligence: First World Conference, xAI
  2023, Lisbon, Portugal, July 26--28, 2023, Proceedings, Part II}.
\newblock Springer Nature.

\bibitem[Mitchell et~al., 2008]{mitchell2008frontal}
Mitchell, D.~J., McNaughton, N., Flanagan, D., and Kirk, I.~J. (2008).
\newblock Frontal-midline theta from the perspective of hippocampal
  “theta”.
\newblock {\em Progress in neurobiology}, 86(3):156--185.

\bibitem[Musaeus et~al., 2018]{musaeus2018}
Musaeus, C.~S., Engedal, K., H{\o}gh, P., Jelic, V., M{\o}rup, M., Naik, M.,
  Oeksengaard, A.-R., Snaedal, J., Wahlund, L.-O., Waldemar, G., et~al. (2018).
\newblock Eeg theta power is an early marker of cognitive decline in dementia
  due to alzheimer’s disease.
\newblock {\em Journal of Alzheimer's Disease}, 64(4):1359--1371.

\bibitem[Nirabi et~al., 2021]{nirabi2021}
Nirabi, A., Abd~Rahman, F., Habaebi, M.~H., Sidek, K.~A., and Yusoff, S.
  (2021).
\newblock Machine learning-based stress level detection from eeg signals.
\newblock In {\em 2021 IEEE 7th International Conference on Smart
  Instrumentation, Measurement and Applications (ICSIMA)}, pages 53--58. IEEE.

\bibitem[{R Core Team}, 2022]{R2022}
{R Core Team} (2022).
\newblock {\em R: A Language and Environment for Statistical Computing}.
\newblock R Foundation for Statistical Computing, Vienna, Austria.

\bibitem[Ribeiro et~al., 2016]{ribeiro2016should}
Ribeiro, M.~T., Singh, S., and Guestrin, C. (2016).
\newblock " why should i trust you?" explaining the predictions of any
  classifier.
\newblock In {\em Proceedings of the 22nd ACM SIGKDD international conference
  on knowledge discovery and data mining}, pages 1135--1144.

\bibitem[Rigby and Stasinopoulos, 2005]{Rigby2005}
Rigby, R.~A. and Stasinopoulos, D.~M. (2005).
\newblock Generalized additive models for location, scale and shape,(with
  discussion).
\newblock {\em Applied Statistics}, 54:507--554.

\bibitem[Roberts et~al., 2013]{roberts2013oscillatory}
Roberts, B.~M., Hsieh, L.-T., and Ranganath, C. (2013).
\newblock Oscillatory activity during maintenance of spatial and temporal
  information in working memory.
\newblock {\em Neuropsychologia}, 51(2):349--357.

\bibitem[Stasinopoulos et~al., 2017]{stasinopoulos2017}
Stasinopoulos, M.~D., Rigby, R.~A., Heller, G.~Z., Voudouris, V., and Bastiani,
  F.~D. (2017).
\newblock {\em Flexible regression and smoothing : using GAMLSS in R}.
\newblock R. Chapman and Hall/CRC.

\bibitem[Tadel et~al., 2011]{tadel2011}
Tadel, F., Baillet, S., Mosher, J.~C., Pantazis, D., and Leahy, R.~M. (2011).
\newblock Brainstorm: a user-friendly application for meg/eeg analysis.
\newblock {\em Computational intelligence and neuroscience}, 2011:1--13.

\bibitem[Tang et~al., 2022]{tang2022}
Tang, C., Li, Y., and Chen, B. (2022).
\newblock Comparison of cross-subject eeg emotion recognition algorithms in the
  bci controlled robot contest in world robot contest 2021.
\newblock {\em Brain Science Advances}, 8(2):142--152.

\bibitem[Thornberry et~al., 2023]{thornberry2023}
Thornberry, C., Caffrey, M., and Commins, S. (2023).
\newblock Neural evidence for navigation efficiency: human theta oscillatory
  power decreases are associated with spatial learning in a virtual water maze
  task.

\bibitem[Vahid et~al., 2018]{vahid2018machine}
Vahid, A., M{\"u}ckschel, M., Neuhaus, A., Stock, A.-K., and Beste, C. (2018).
\newblock Machine learning provides novel neurophysiological features that
  predict performance to inhibit automated responses.
\newblock {\em Scientific reports}, 8(1):16235.

\bibitem[Visser and Speekenbrink, 2010]{Visser2010}
Visser, I. and Speekenbrink, M. (2010).
\newblock {depmixS4}: An {R} package for hidden markov models.
\newblock {\em Journal of Statistical Software}, 36(7):1--21.

\bibitem[Zucchini and MacDonald, 2016]{zucchini2016}
Zucchini, W. and MacDonald, I.~L. (2016).
\newblock {\em Hidden Markov models for time series: an introduction using R}.
\newblock Chapman and Hall/CRC, second edition.

\bibitem[{\.Z}ygierewicz et~al., 2022]{zygierewicz2022decoding}
{\.Z}ygierewicz, J., Janik, R.~A., Podolak, I.~T., Drozd, A., Malinowska, U.,
  Poziomska, M., Wojciechowski, J., Ogniewski, P., Niedbalski, P., Terczynska,
  I., et~al. (2022).
\newblock Decoding working memory-related information from repeated
  psychophysiological eeg experiments using convolutional and contrastive
  neural networks.
\newblock {\em Journal of Neural Engineering}, 19(4):046053.

\end{thebibliography}

\end{document}